\documentstyle[twoside,fleqn,espcrc2,epsf]{article}

\epsfverbosetrue
\def\Psb{\overline{\Psi}}
\def\ob{\overline{\omega}}
\def\xib{\overline{\xi}}
\def\Ds{D\!\!\!\!/\,}
\def\ds{\partial\!\!\!\!/\,}
\def\dg{\dagger}

\newcommand{\AmS}{{\protect\the\textfont2
  A\kern-.1667em\lower.5ex\hbox{M}\kern-.125emS}}

\title{Chiral gauge theories with domain wall fermions}

\author{Maarten Golterman,\address{Department of Physics,
        Washington University\\
        St. Louis, MO 63130-4899, USA}%
        \thanks{speaker at conference}
	Karl Jansen,\address{Department of Physics,
	University of California at San Diego\\
	La Jolla, CA 92093,  USA}%
	\thanks{present address:
                DESY, Notkestrasse 85, 22603 Hamburg 52, Germany}
	Don Petcher,$^{\rm a}$
	\thanks{present address:
                Covenant College, Lookout Mountain, GA 30750, USA}
        and 
	Jeroen C. Vink$^{\rm b}$}
       
\begin{document}

\begin{abstract}
We have investigated a proposal to construct chiral gauge theories on
the lattice using domain wall fermions.  The model contains two opposite
chirality zeromodes, which live on two domain walls.  We couple only one
of them to a gauge field, but find that mirror fermions which also couple
to the gauge field always seem to exist.
\end{abstract}

\maketitle

\section{Model}

A very elegant mechanism for obtaining lattice chiral fermions was
recently proposed \cite{kaplan}.  The idea is to start from Wilson fermions
in $d+1$ dimensions, where $d$ is the dimension of space-time, and give these
fermions a mass $m_0$ of the order of the cutoff.  
However, a domain wall like defect
is introduced by making the mass term dependent on the extra dimension, 
choosing $m^s=-m_0$ for $s<0$ and $m^s=m_0$ for $s>0$ ($s$ labels the
coordinates of the extra dimension).  It was shown that a massless mode with
positive chirality exists in this model \cite{kaplan}, which is exponentially
bound to the defect at $s=0$, which is identified with $d$ dimensional 
space-time.

In a finite volume an antidomain wall exists due to the (anti)periodic
boundary conditions, and consequently, a negative chirality mode, bound
to this defect, exists (other choices of boundary conditions are possible, 
but do not alter the conclusions \cite{shamir}). 
One can study this system coupled to external smooth gauge fields, and one
finds that any anomalous charge created on the domain wall by this gauge field
is carried off through a Goldstone-Wilczek current to the antidomain wall
\cite{kaplan,jansen,goljankap}.

In this work, we study the coupling to dynamical gauge fields. 
A detailed account can be found in ref. \cite{goletal}, to which we also
refer the reader for notations.   We 
introduce a waveguide $WG=\{s:s_0\le s\le s'_0\}$, with $s_0$ on one side, and 
$s'_0$ on the other side of the domain wall.  Only within the waveguide, the
fermions are coupled to a dynamical gauge field $U_\mu$.  This breaks gauge 
invariance in the fermion hopping terms across the waveguide boundary,
which is restored by promoting these terms to Yukawa couplings with a scalar
field $V$, which takes values in the gauge group $G$ \cite{kaplat92}.
For another approach with $s$-independent gauge fields, see ref. \cite{nn1}.
The model is defined by the action
\begin{eqnarray}
 S_\Psi & = & \sum_{s\in WG} \Psb^s(\Ds(U) -W(U) + m^s)\Psi^s \nonumber \\
& + & \sum_{s\not\in WG}  \Psb^s(\ds-w+ m^s)\Psi^s 
  +   \sum_s \Psb^s\Psi^s \nonumber\\
& - & \sum_{s\not=s_0-1,s_0'}[\Psb^sP_L\Psi^{s+1}
                         + \Psb^{s+1}P_R\Psi^s ] \nonumber \\
& - & y(\Psb^{s_0-1}VP_L\Psi^{s_0}  
                 +  \Psb^{s_0}V^{\dg}P_R\Psi^{s_0-1}) \nonumber \\
& - & y(\Psb^{s_0'}V^{\dg}P_L\Psi^{s_0'+1} + \Psb^{s_0'+1}VP_R\Psi^{s_0'}),
\end{eqnarray}
where $m^s$ is the $s$-dependent mass.

For $y=0$ the regions inside and outside the waveguide decouple, and
new zeromodes appear on one of the waveguide boundaries, with negative
chirality just inside, and with positive chirality just outside the
waveguide \cite{goletal,shamir}.  (They appear only at one of the boundaries:
which one depends on details.)
 The inside mode couples to the gauge
field, rendering the theory vectorlike.  The key question now is whether
this situation pertains for all values of the Yukawa coupling $y$, or that
a PMS phase exists for large values of $y$
\cite{PMS}, where the fermion spectrum
around the waveguide boundary does not contain light fermion modes.  This
phenomenon takes place in a number of Higgs-Yukawa models, and if this would
be the situation in the case at hand, 
the unwanted modes at the waveguide boundary would decouple.
We emphasize that
there are no simple anomaly arguments to rule out this possibility
\cite{goletal}.

\section{Effective model}

We have studied this question for $U_\mu=1$, as it is the fermion-scalar
dynamics that would be responsible for the existence of a PMS phase.
This is reasonable, as $U_\mu$ (in the Landau gauge) 
should be close to one in an asymptotically
free theory.  At $y=0$ the action can be diagonalized, and one finds
two massless Dirac modes, $\omega=\omega_R+\omega_L$ and $\xi=\xi_R+\xi_L$,
where $\omega_{R(L)}$ is the right(left)-handed mode at the (anti)domain wall,
and $\xi_{L(R)}$ is the left(right)-handed mode just inside (outside) the
waveguide boundary.  These modes all have a chiral spectrum for momenta
in the region 
$\frac{1}{2}{\hat p}^2=\sum_\mu(1-\cos{p_\mu})<\frac{1}{2}p_c^2
\equiv |2-m_0|$ 
around $p=0$ \cite{kaplan,janschm}.
Modes with momenta outside this region are delocalized in $s$, and their
wavefunctions are negligible at the domain walls and waveguide boundary.
This includes all species doubler momenta $p=(\pi,0,\dots)$ {\it etc}.
In writing an effective action for the light modes, we may therefore
ignore these large $p$ modes.  We will also ignore the Yukawa couplings
of the domain wall modes ($\omega_{L,R}$), since their wavefunctions are
exponentially suppressed at the waveguide boundaries.  This leads us to an
effective model for $\omega$ and $\xi$:

\begin{eqnarray}
 S_{\rm eff} &\! = \!& \sum_{|{\hat p}|<p_c}\Bigl(i\ob_p\gamma_\mu\sin{p_\mu}
 \omega_p 
 + i\xib_p \gamma_\mu\sin{p_\mu}\xi_p\Bigr) \nonumber \\
 &\! +\!& y\sum_{|{\hat p}|,|{\hat q}|<p_c}\xib_p\Bigl(V^\dg_{q-p}P_L
+ V_{p-q}P_R\Bigr)\xi_p.
\end{eqnarray}
 
This form of the effective action is reminiscent of 
a hypercubic Higgs-Yukawa model \cite{hyper}, 
in that in both models the Yukawa couplings are 
suppressed for large momenta.  This leads to the suspicion that no PMS
phase will exist in our model, as no such phase exist in the hypercubic
model.

\section{Numerical results}

The obvious thing for a numerical investigation, in order to find out whether
a PMS phase exists or not, is to compute fermion masses.  
However, in
particular for large $y$ and small scalar hopping parameter $\kappa$
 (the region
of interest \cite{goletal}), 
the propagators become very small and noisy due to the 
strongly fluctuating scalar field, and we have not been able to compute
reliable fermion propagators in that region (we mostly used the quenched
approximation).  We studied the eigenvalue spectrum of the Dirac operator
for the effective action, eq. (2), for $G=U(1)$ and $d=2$, and compared
it with $d=2$ Higgs-Yukawa models which are known to have or lack a PMS
phase.  The distribution of eigenvalues in the complex plane is a very good
measure of the existence of the PMS phase \cite{baretal}.  Results are
shown in fig. 1, indicating that indeed no PMS phase exists.

\begin{figure}[htb]
 \centerline{ \epsfysize= 5cm  \epsfbox{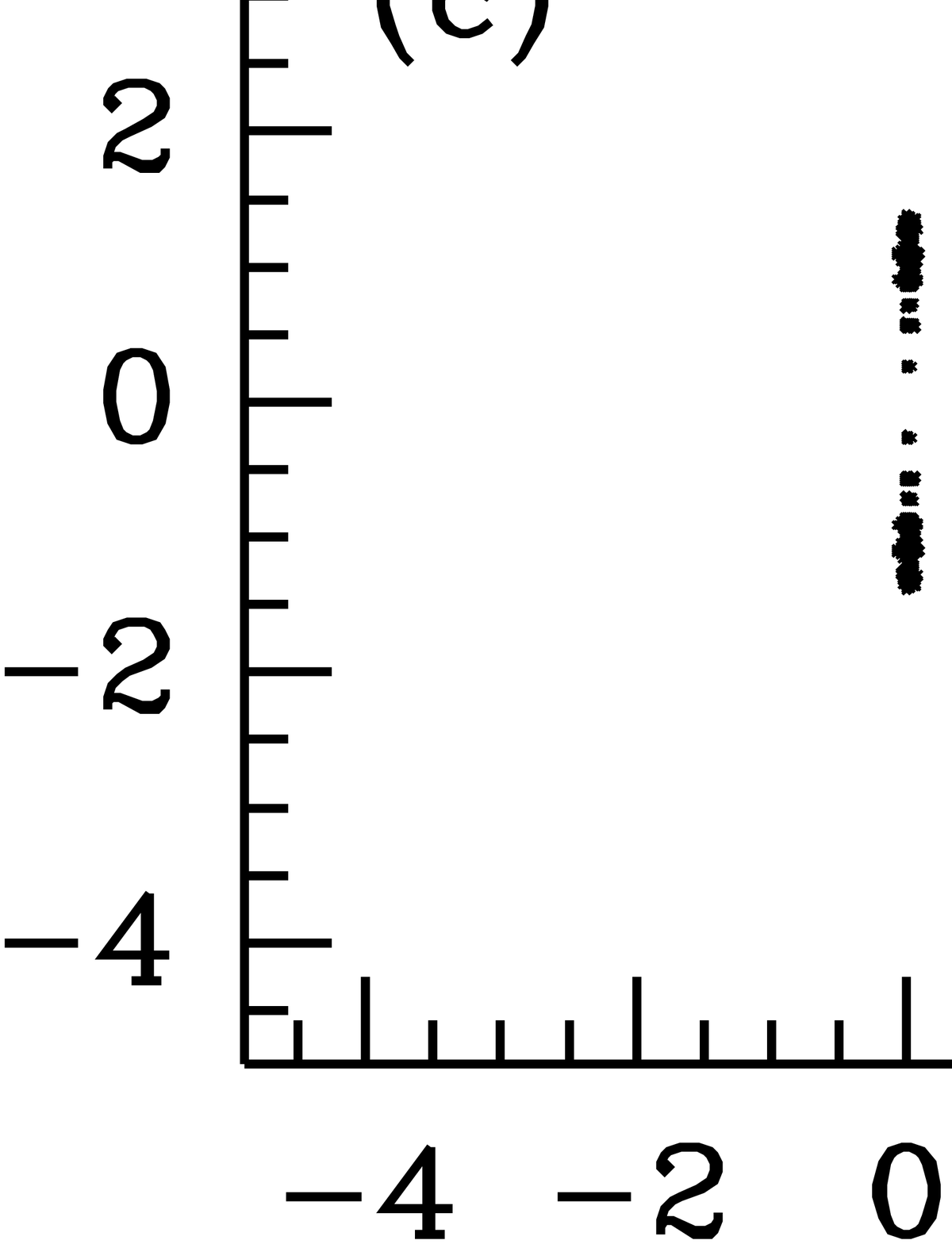} 
 }
 \caption{
Eigenvalue spectra for the effective model (figs. a),
 and for two reference models with (figs. b) and without (figs. c) a PMS
 phase.
 The left, middle and right figures are for $y=0.2$, $1.0$
 and $4.0$ resp.  The lattice size is $L^2=12^2$ and $\kappa=0.1$.}
\vspace{-5mm}
\end{figure}

We have also computed the $\xi$ mass in the broken phase ({\it i.e.}
for large $\kappa$) of the full model (eq. (1)), 
where the signal to noise ratio is better, again for $U(1)$ and $d=2$.
(For a discussion of the finite volume $d=2$ scalar phase structure,
see ref. \cite{goletal}.)  Results for the $\kappa$ dependence are
shown in fig. 2 for strong Yukawa coupling $y=2$, 
which are consistent with the weak coupling behavior
$m_F\approx yv$ for $\kappa\searrow\kappa_c\approx 0.5$.  Near a FM-PMS
phase transition one would have expected a fermion mass increasing with
$\kappa\searrow\kappa_c$ \cite{PMS}.  
Again, this indicates that no PMS phase exists
in our model.

\begin{figure}[htb]
\centerline{ \epsfysize = 4cm \epsfbox{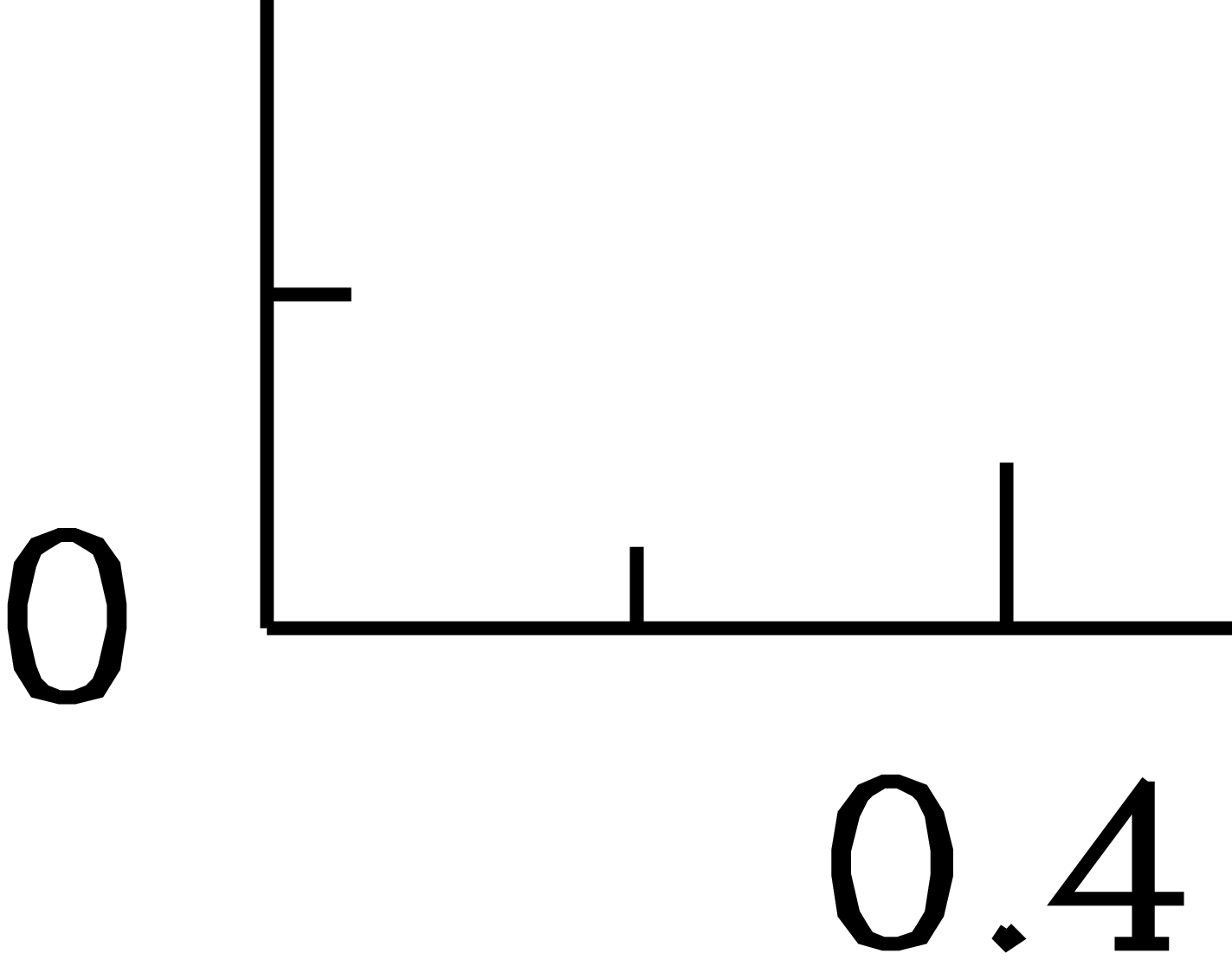}} 
 \caption{
$\kappa$ dependence of the waveguide fermion mass at $y=2$
on a $12^226$ lattice.}
\vspace{-5mm}
\end{figure}

\section{Final remarks}

We conclude with some remarks, most of
which are more extensively discussed in
ref. \cite{goletal}.

$\bullet$ We have tried unquenched computations, which resulted
in very poor statistics, and very low acceptance rate of the hybrid MC
algorithm.  Unquenched results were not inconsistent with our quenched
computations.

$\bullet$ We found good agreement between fermion masses computed
in the full and effective models.

$\bullet$ The scalar field is nothing else than the longitudinal
gauge field, and it is due to the fluctuations of this field that the
model produces mirror fermions, rendering the model vectorlike: both
$\omega_R$ and $\xi_L$ couple to the gauge field.

$\bullet$ A crucial role is played by the ``effective momentum
cutoff" $p_c$, which is critical in removing the doublers.  However,
from eq. (2) it is also clear that this cutoff plays an important role
in the (non)existence of a PMS phase.

$\bullet$ We have also considered
a staggered fermion formulation of the domain wall approach, based on
the notion that in that case no $p_c$ appears.  However, in this case,
the flavor structure causes similar problems \cite{golvin}.

$\bullet$ The model considered here is more general than the
original way of coupling to dynamical
gauge fields proposed in ref. \cite{kaplan}.  

$\bullet$ There exists a proposal to keep the volume in the extra
dimension strictly infinite \cite{nn1,nn}, based on the idea that in that way
the zeromode on the antidomain wall is avoided from the start.  
Ref. \cite{shamir2} discusses a possible relationship between this approach 
and the one reported on here.

\end{document}